%% file: content.tex
\documentclass[
  10pt, 
  a4paper, 
  oneside, 
  headinclude,footinclude, 
  BCOR5mm, 
]{scrartcl}

\input{structure.tex}

\usepackage{setspace}
\usepackage{graphicx}
\usepackage{float}

\begin{document}

\begin{center}
{\large \textbf{Pattern Matching and Classification of Clusters in Collision Cascades}}\\
  
  Utkarsh Bhardwaj$^a$, Andrea E. Sand$^b$, Manoj Warrier$^{ac}$ \\
  {\small
    \textit{
      $^a$Computational Analysis Division, BARC, Visakhapatnam, Andhra Pradesh,
      India - 530012 \\
      $^b$Department of Physics, P.O. Box 43, FI-00014 University of Helsinki,
      Finland \\
      $^c$Homi Bhabha National Institute, Anushaktinagar, Mumbai, Maharashtra,
      India - 400 094 \\
    }
  }
\end{center}

\section*{Abstract}
The structure of defect clusters formed in a displacement cascade plays a
significant role in the micro-structural evolution during irradiation. Molecular
dynamics simulations have been widely used to study collision cascades and
subsequent clustering of defects. We present a novel method to
pattern match and classify defect clusters. A cluster is characterized by the
geometrical and topological histograms of its angles and distances which can
then be used as similarity metrics. The technique is demonstrated by matching
similar clusters for different cluster shapes like ring, crowdions etc. in a
database of cascade damage configurations in Fe and W at different energies. 
We further use graph based dimensionality
reduction techniques and unsupervised machine learning on the features of all
the clusters present in the database to find classes of clusters. The
classification successfully separates out many already known categories of
clusters such as crowdions, planar crowdion pairs, rings and perpendicular
crowdions. The dimensionality and size of different classes provides a broad
categorization of classes. The distribution of different classes of shapes
among cascades of different elements and energies shows the exclusivity of
shapes to elements and energies. We discuss the key points and computational
efficiency of the algorithms along with the various prominent results of
their application. We discuss the motivation for using machine learning and
statistics for the problems and compare different techniques. The
algorithms along with the supporting analysis and visualizations give an
unsupervised approach for classification and study of defect clusters in
cascades. The distribution of cluster shapes and structures along with the
shape properties like diffusivity, stability, etc. can be used as input to
higher scale models in a multi-scale radiation damage study.

\section{Introduction}

The defects formed during the displacement cascades due to irradiation are the
primary source of radiation damage \cite{Stoller1997, Stoller2012,
KMCIrrDamWirth, WirthCuFe, BecquartMDKMC, StollerMDKMC, KaiBjorkas}. The
defects in metals with body-centered cubic structure are produced in the form
of single point defects (interstitials and vacancies) or clusters of such
defects. The point defects and glissile clusters diffuse after the cascade to
either annihilate or form bigger defect clusters. The structural details of
primary point defect clusters (formed as a direct consequence of the cascade)
define the diffusion, recombination, thermal stability and their other
characteristics \cite{Stoller2012, OSETSKY200185, GAO2000213, BACON20001} which
in the long term determine the micro-structural changes in the material
\cite{SINGH1997107, GOLUBOV200078, OSETSKY200065, BECQUART200639,
OSETSKY2002852}. These properties have an affect on the results of higher scale
models like Monte Carlo methods, rate theories etc. \cite{OSETSKY200065,
BECQUART200639, KaiBjorkas, Cas17}. The glissile clusters can move and interact
with other defects and grain boundaries whereas the sessile clusters can be
nucleation centers for defect-growth. The interaction of these clusters with
other defects will decide the micro-structural changes due to irradiation.
Classification and taxonomy of all possible clusters in different irradiated
samples is the first step in the systematic study of properties of clusters
and their effects.

We approach the problem of characterization, matching and classification of
point defect clusters using the techniques of topology, geometric shape pattern
matching, statistics and unsupervised machine learning. A cluster is
characterized by the normalized histogram of angles between the neighbouring
defect triads and pair wise distances. We use this histogram as a feature
vector for the cluster. To find the similarity between two clusters it is
sufficient to find the distance between the two feature vectors that represent
them. The histograms of angles, distances and adjacency are a simple, deformation
invariant, computationally efficient but powerful way to characterize a shape
\cite{shape1}. Other advanced topological techniques such as barcode like
representation from simplices \cite{MATE20141180} can also be effective. We
further use topological network graph based dimensionality reduction techniques
on the feature vectors. The network graph based dimensionality reduction
techniques are well established ways to find a representation in the reduced
dimensional space such that the distance between the distinct points is
maximized \cite{tsne, umap} and similar points are represented closely. The
reduced dimensional representation is also known as embedding of the feature
representation. We then use an unsupervised clustering algorithm to classify the
clusters without any inputs and assumptions made. With the help of
classification we can explore the dimensionality, sizes and exclusivity of
certain shapes in elements and energy ranges. Some classes of clusters we found
in Fe have been reported in \cite{DEZERALD2014219}.

We present the results on a database of cascade damage configurations simulated
by molecular dynamics (MD) \cite{andrea, nordlund2018improving}. The database
has Fe and W cascades at different energies ranging from 10 keV to 200 keV. The
code is integrated with interactive visualizations built using webGL interface
making it easy to qualitatively study and understand the results.

Section 2 describes the various algorithms and methods used in every step with
a brief discussion on the suitability of the method used to solve the problem
and comparison with related techniques. The implementation of the methods along
with interactive web-app can be found at \url{https://github.com/haptork/csaransh}
open source repository. Section 3 presents results of the faster defects
identification algorithm implementation, pattern matching and searching of
similar clusters in the database, dimensionality reduction and classification
of clusters for the database. The results for the database can be accessed via
interactive web application available at \url{https://haptork.github.io/csaransh}.
Finally, we conclude by discussing the significance of using the presented
method for the study of radiation damage.


\section{Algorithms and Methods\label{impl}}

The problems of classification and pattern matching find state-of-the-art
solutions in machine learning and artificial intelligence. After finding
the defect clusters, we divide the problem of characterization, matching and
classification of clusters into the following machine learning template
steps:

\begin{enumerate}
  \item Feature representation
  \item Dimensionality reduction
  \item Classification with an unsupervised clustering algorithm
\end{enumerate}

In the feature representation step, we find a representation of a cluster that
approximately characterizes its shape and ensures translational, scale and
rotational invariance. The distance between the two feature representations can
be used as the (dis-)similarity metric of their clusters. The next step viz.
dimensionality reduction, transforms the feature space into a space with
reduced principle dimensions. The network graph based dimensionality reduction
approaches that we employ aim to increase the probability that similar points
appear nearer and dissimilar points have distant embedding (feature vector in
new space). By reducing to two or three dimensional space we can qualitatively
check the global and local relationships between all the clusters and also
explore how the embeddings separate out known groups like interstitials and
vacancy clusters, different sizes etc.  Next, we use the embeddings for
classification using unsupervised clustering.  We use density
based clustering algorithms, that group together the points (here, cluster
embeddings) that are densely packed together and ignore the noise. We
then explore different statistical properties of the classes like
dimensionality, size and specificity of shapes among different elements and
energies. 

Starting from how we find the defects and clusters from MD simulation output to
classification, we provide a detailed description of each step in the following
sub-sections.

\subsection{Finding Defects and Clusters from MD simulations\label{idalgo}}

To find the defects from all the atoms in the last frame of
an MD simulation, we use an algorithm that is similar to Wigner-Seitz 
method \cite{gibson1960dynamics, nordlund1997point}, but more efficient and
also finds extra interstitial-vacancy pairs crucial to defining shape of the
cluster e.g. a crowdion cluster can be formed of three interstitials and two
vacancies; although the defect count is one interstitial but the shape and
dependent properties are a function of all the five defects. 

The algorithm can be divided into two stages. The first stage is to assign closest
lattice site to every atomic coordinate that is achieved by using the modulo
arithmetic \cite{BHARDWAJ2016263}. The steps are as follows:

\begin{itemize}
  \item For every atom:
  \begin{enumerate}
    \item Find the modular position by taking modulo (remainder) of coordinate
values by lattice constant. This gives the position of the atom within a unit cell.
    \item Using the modular position, find the minimum distance lattice site position
      in the first unit-cell out of all the lattice sites possible for the
      given lattice structure (bcc, fcc etc.).
    \item Find the unit cell in which the atom lies by taking quotient of coordinate
values by lattice constant.
    \item Using the closest lattice site in the first unit cell found in the second
step and the unit cell in which the atom lies found in the third step, assign
the closest lattice site position to the atom.
  \end{enumerate}
\end{itemize}

The algorithm can run in parallel for all the atoms. The implementation utilizes
the crystal structure information to calculate the closest lattice site using
modulo arithmetic. The other implementations that use the initial atomic
coordinates as template require extra memory to store the template structure.
They either require simple but slow $ \mathcal{O}( N^2) $ look-up or need to
build a special datastructure for nearest neighbour search such as k-d tree that
is $ \mathcal{O}( kN log N) $ to build and $ \mathcal{O}( N log N) $ for the
look-up of neighbours for all the N atoms. Our implementation takes only single
pass through the data making the operation $ \mathcal{O} (N) $. It is also
completely parallelizable. Since it is just a single, continuous pass through
the data, it gives good cache performance. However, a template free approach is
only possible for the crystal lattices, where points can be enumerated and one
to one mapping can be implemented between the enumeration and the lattice site
coordinates in the crystal. In the stage two we make use of the mapping to go
from index of a lattice site to the atomic coordinates and vice-verse. The steps
are as follows:

\begin{enumerate}
  \item Initialize a boolean array with false values having size same as total
    number of atomic sites.
  \item Iterate over the array of atomic coordinates:
  \begin{enumerate}
    \item Check the value of the boolean array at index equals to the closest
      site lattice enumeration added for the current atomic coordinate in stage 1.
    \item If the value is false, mark it as true.
    \item Else (if the value is true, implying there was another co-ordinate
      that had this index as closest lattice site), mark the current atomic
      coordinate as interstitial.
  \end{enumerate}
\item Iterate over the boolean array and mark all the indices (or the
  coordinates corresponding to the indices) that hold false value (implying
  that no atom had these indices as closest lattice site) as vacancy.
  \end{enumerate}

To make the algorithm output extra interstitial-vacancy pairs, we can change
the algorithm such that in the case where the value at index in the boolean
array is already true, mark not just the current atom as interstitial, but also
mark as interstitial the first atom that made the boolean array value true and the
lattice site for the index as vacancy. We can keep an extra flag to mark these
as pseudo defects.

Unlike template based Wigner-Seitz or similar methods, the algorithm never
holds the whole perfect lattice in memory and does not require neighbour
search, runs faster with $ \mathcal{O}( N ) $ time-complexity and uses simpler
cache-friendly datastructures.

The machine learning method described in \cite{BHARDWAJ2016263, BUKKURU2017258}
that uses max-space clustering on offsets of atoms, has $ \mathcal{O}( N log N) $ 
time-complexity but probably has faster implementation than k-d tree, requires
no lattice or any other information. However, it can only be used to identify
interstitials and not the vacancies. The other sphere based methods
\cite{Stoller2012, caturla1996ion}, require a threshold and would also require
a neighbour search for finding interstitials and vacancies. The other
implementation of the threshold based sphere method as given in
\cite{WARRIER2015457} requires initial position with atom i.d. to match between
initial and final positions of all the atoms.

The algorithm presented in this work takes only final positions and crystal
structure as input, makes no extra assumptions and on a single run can output
both interstitials, vacancies as well as structural defects that can be 
separately labelled, e.g. as pseudo-defects. These pseudo-defects do not appear
in an analysis based on the use of Wigner-Seitz cells.

To find the clusters from the identified defects, we use distance threshold and
union-find \cite{UF} data structure to find the clusters efficiently
\cite{WARRIER2015457}. The extra interstitials and vacancies found by the method
also contribute to the clustering so that all the defects in crowdions, complex
ring structures etc. are grouped in the same cluster.

\subsection{Finding the Feature vector for a cluster}

We characterize a cluster by histograms of: angles of triplets, distances
between pairs and number of neighbours of the defects in the cluster. The first
two features are based on geometry while the last one is based on topology.
Following are the algorithms for each.

\subsubsection{Histogram of Angles}

\begin{enumerate}
  \item Choose a bin size for the angles between 0 and 180, a reasonable default can
    be between 3 to 15 degrees. 
  \item Initialize all the bins with zeroes.
  \item For each vacancy in the cluster find the angle that two neighbouring
    interstitials or two neighbouring vacancies subtend. Find the bin where the
    angle falls and increase the count of that bin by one.
 \end{enumerate}

 We find angles only with vacancy as the pivot of the triad and the other two defects
 are either both interstitials or both vacancies. Since, vacancies are at the fixed
 lattice points, it is equivalent to measuring the arrangement of interstitials and
 other vacancies that are around. Adding more angles such as with interstitials as
 pivot adds no extra information about the shape.

\subsubsection{Histogram of Distances}
\begin{enumerate}
  \item Choose a bin size for normalized distances between 0.0 to 1.0, a reasonable default can be
    between 0.01 to 0.1.
  \item Initialize all the bins with zeroes.
  \item For each vacancy in the cluster find the distances from its neighbours. 
  \item Divide all the distances by the maximum distance, to normalize between
    0.0 and 1.0.
  \item Find the bin where distances fall and for each one increase the count
        of that bin by one.  
\end{enumerate}

Again, we take only the distances from fixed vacancy points, since adding more
distances does not further enrich the information.

\subsubsection{Histogram of Neighbours}
\begin{enumerate}
  \item Initialize all the bins with zeroes.
  \item For each defect in the cluster find the number of neighbours and increment by one the
     bin at the index equal to number of neighbours.
\end{enumerate}

For each of the above algorithms we use twice of 4-NN (nearest neighbour) as
the neighbourhood distance that is similar to considering defects in adjacent
unit-cells alone.  We use 5 degrees for the angles bin size and 0.02 as the
distance bin size for the histogram feature representation. However, the results
are not very sensitive to these values. 

The histograms accumulate the local geometry of each defect.
Since the adjacency histogram captures a subset of details captured by the
distance histogram, we drop adjacency features from the further analysis.

There are similar shape histogram features that have been used before to study
molecules \cite{shape1} which involve angles and distances from the center of
mass. However, being independent of the center of mass, the features we use
are more robust to noise and can characterize local structures better e.g. if
there are two structures, one with ring and one with ring and a tail, the
centroid based features will be entirely different while the features presented
above will have same values for the bins that correspond to the ring, although the
latter structure with a tail will have some more additions on the bins
characterizing the tail.

The topology feature we use is simple while there is
still scope to use more advanced features like barcode like representation as
defined in \cite{MATE20141180}. One can also use Hausdorff distance etc. after
getting the defect coordinates of each cluster in their principle axes by
singular value decomposition. However, matching with barcodes or direct
matching after transformation in principle axes and strict distance measures
will not be as efficient as histogram features matching. Also, the direct matching
of structures might suffer because of different orientations, slight distortions
and other noises such as thermal displacements due to residual heat, different
number of total defects in two similar structures etc. Also, the size of
clusters if counted with thermally displaced pseudo defects can grow to a few
hundred point defects. For these reasons we find the histogram based features
simple and efficient for our case as shown in the results section.

For a cluster, we can find other similar clusters by comparing the distance
between histograms. We can use different distance metrics such as chi-square,
correlation, Quadratic Form Distance Functions \cite{qfd}. The definitions of
these metrics can be different, however they all consider each bin of the
histogram as a dimension. The euclidean distance between two $n$ dimensional
histograms ($a$ and $b$) can be defined as $\sqrt{\Sigma_{i=1}^n (a_i -
b_i)^2}$.

\subsection{Dimensionality Reduction of the feature vector}

Graph based dimensionality reduction techniques like t-SNE \cite{tsne}
(t-Distributed Stochastic Neighbor Embedding) and UMAP \cite{umap} (Uniform
Manifold Approximation and Projection) are well established techniques for
qualitative analysis of relationships between the data points (here clusters).
By transforming the basis vectors such that the probability of similar points
to appear together increases, these techniques can help in visualizing and
exploring classification and correlation of cluster shapes with other
quantities like dimensionality of the clusters, size of the cluster etc. 
In our case both UMAP and t-SNE give very similar global structure. 
The degree of local spread produced by UMAP is more amenable to the HDBSCAN
(Hierarchical Density-based Spatial Clustering of Applications with Noise)
\cite{hdbscan, hdbscan2} classification algorithm. UMAP is also a more
efficient algorithm than t-SNE. However, t-SNE gives a relatively more even
spread as shown in the results.

\subsection{Classification of all the clusters}

To get an idea about the different shapes of defect clusters present in the
database and arrange them into classes we use unsupervised clustering.
Unsupervised clustering can be used to group together similar data points
without any need of already labelled training data. We use density based
clustering algorithms viz. DBSCAN (Density-based Spatial Clustering of 
Applications with Noise) \cite{ester1996density} with t-SNE and
HDBSCAN with UMAP because HDBSCAN works well with UMAP embeddings and not with
t-SNE. Density based algorithms classify separated similarly dense point regions
into different classes and less dense isolated points as noise. HDBSCAN has the
advantage that it only depends on a single integer parameter as opposed to two
in DBSCAN. We can use other popular unsupervised classification algorithms like
K-Means but unlike density based algorithms K-Means does not ignore the noise
in the data and requires number of classes as user input i.e. a-priori
information about the number of distinct cluster classes need to be known. Both
the density based algorithms used give very similar classifications as shown in
the results section.


\section{Results}

This section discusses the results of the methods described above applied to a
collection of defect configurations from collision cascades in bulk W and Fe.
Cascades in W were simulated using the method described in \cite{andrea}, with
the interatomic potential by Derlet et al. \cite{Derlet}, stiffened by
Bj{\"o}rkas et al. \cite{Bjo09}. Cascades in Fe were simulated with the same
method, using the Ackland-Mendelev interatomic potential \cite{Ack04}. All
cascades were simulated with an initial temperature at 0 Kelvin, and with
electronic stopping applied to atoms with energies above 10 eV \cite{San15},
and evolved for 40 ps. The database contains total 76 collision cascades out of
which, 36 are of W at energies 100 keV, 150 keV, 200 keV and 40 are of Fe at
energies 10 keV, 20 keV, 50 keV, 100 keV, 150 keV. The cascades have more than a
thousand vacancies and interstitial clusters of various shapes and sizes,
providing a varied database to work with. The following subsections discuss the
significance and validity of each method using the results found on the
database.

\subsection{Defects and Clusters from Cascades}

\begin{figure}[H]
  \centerline{\includegraphics[width=.9\linewidth]{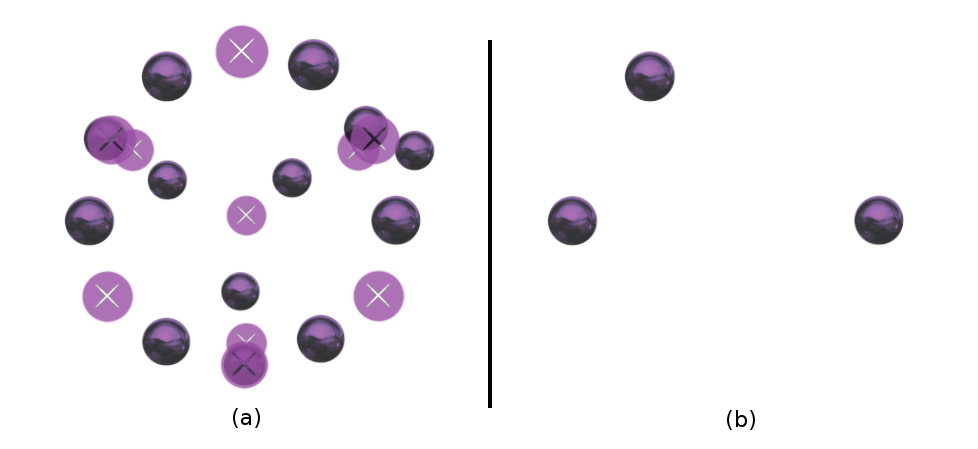}}
  \caption{\label{wrwo}
    A cluster in W cascade at 200 keV (a) with and (b) without extra
    displaced atoms and their lattice sites. Solid spheres are interstitials
    and flat-cross glyph represents vacancies. The ring like shape of the cluster
    is evident only in (a).
}
\end{figure}

The algorithm presented to identify the defects in Section \ref{idalgo} can
find out the extra interstitial-vacancy pairs crucial for defining the shape of
a cluster. Fig. \ref{wrwo} shows a cluster of W cascade at 200 keV with and
without these extra interstitial-vacancy pairs. The ring shape of the cluster
can only be identified with the extra pairs included.

We also validate the accuracy of the algorithm by matching the number of
defects against that found with the template based Wigner-Seitz method implemented
in Ovito \cite{ovito}. The results match for all the cascades in the database.

\subsection{Feature Vector for a cluster}

\begin{figure}[H]
  \centerline{\includegraphics[width=.9\linewidth]{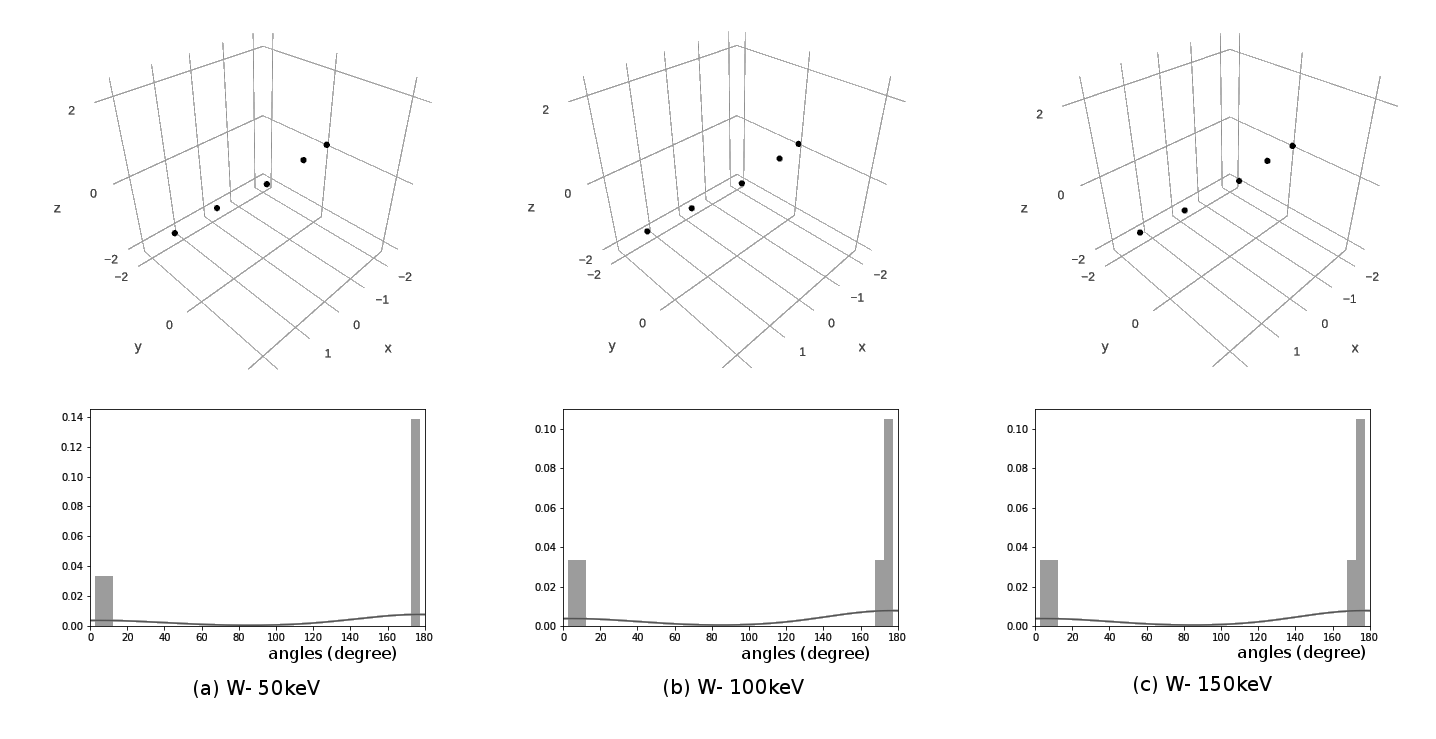}}
  \caption{\label{linear}
    Linear crowdion clusters along with their triad angle histogram
    feature. The crowdions in (b) and (c) are the results of a search 
    for closest feature to (a) in the database. The histogram shows
    high number of occurrences in 0 degrees and 180 degrees that is
    also qualitatively a characteristic for a linear shape.
}
\end{figure}

\begin{figure}[H]
  \centerline{\includegraphics[width=.9\linewidth]{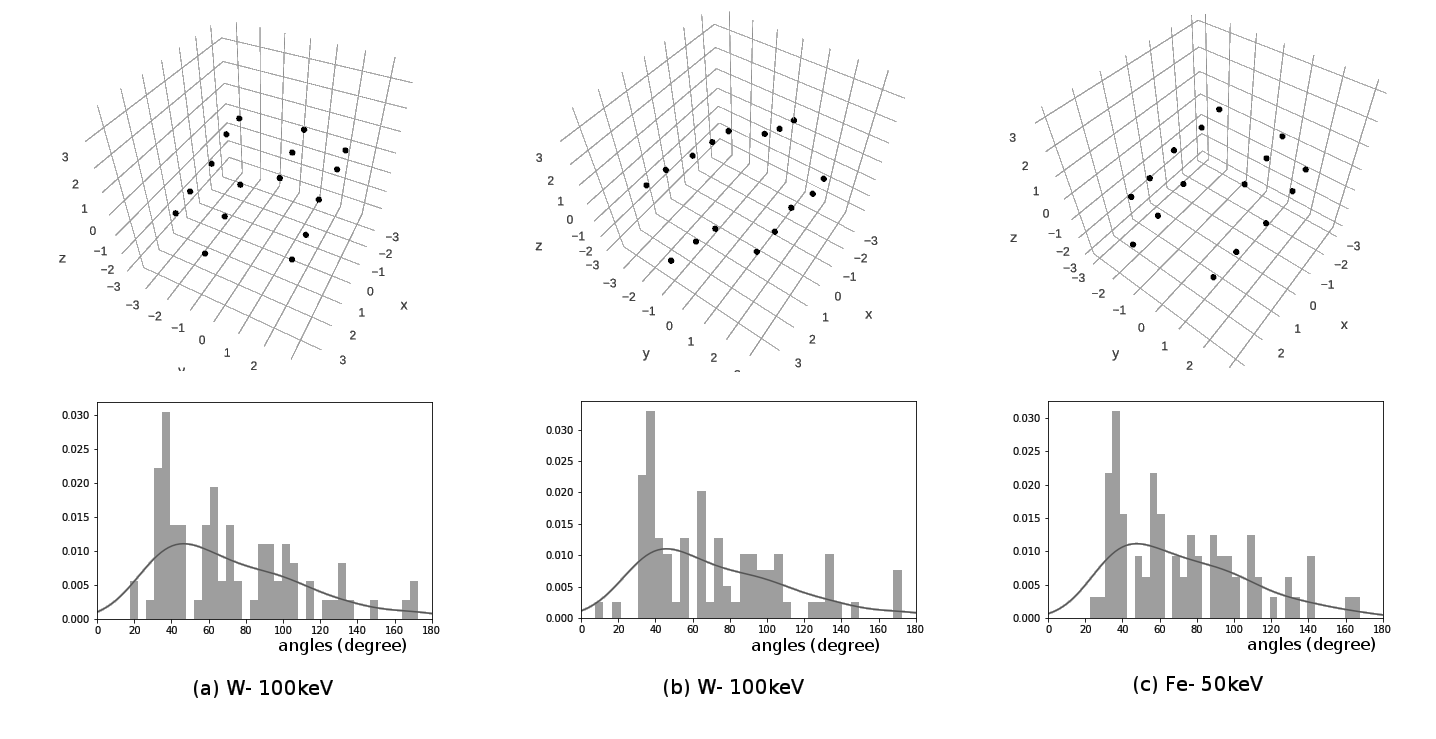}}
  \caption{\label{stacked-crowdion}
    Four stacked crowdions and dumbbells along with their triad angle histogram
    feature. The clusters in (b) and (c) are the results of a search 
    for closest feature to (a) in the database. The histogram is a
    characteristic for the shape.
}
\end{figure}

\begin{figure}[H]
  \centerline{\includegraphics[width=.9\linewidth]{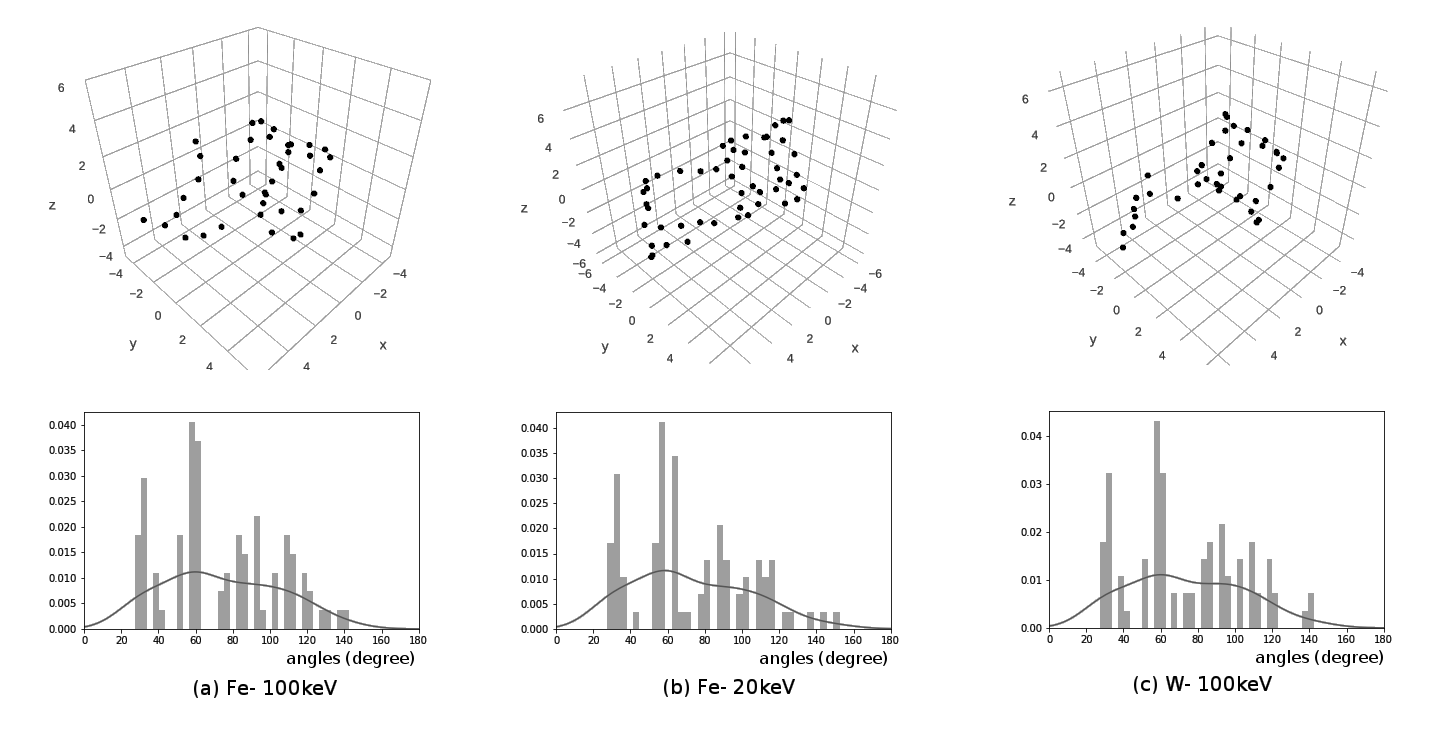}}
  \caption{\label{ring}
    Ring like clusters along with their triad angle histogram
    feature. The rings in (b) and (c) are the results of a search 
    for closest feature to (a) in the database. The histogram is a
    characteristic for ring shape.
}
\end{figure}

\begin{figure}[H]
  \centerline{\includegraphics[width=.9\linewidth]{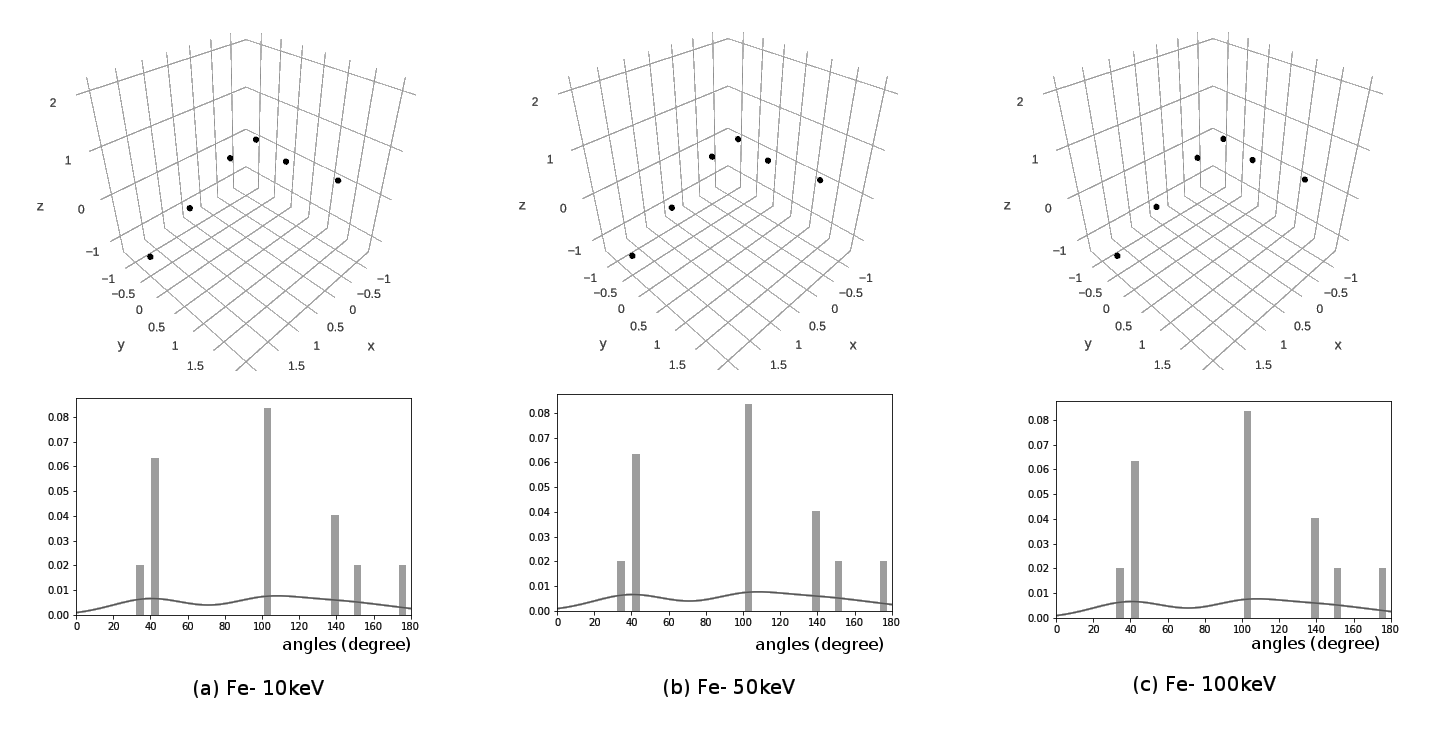}}
  \caption{\label{tee}
    Tee shaped non-planar pair of dumbbells along with their triad angle
    histogram feature. The rings in (b) and (c) are the results of a search for
    closest feature to (a) in the database. The histogram is a characteristic
    for the shape.
}
\end{figure}

The histogram features retain the information about the characteristic shape of a
cluster that are invariant to rotation, scale etc. while ignoring the
superfluous details. Fig. \ref{linear} to Fig.  \ref{tee}
show the triad angle histograms next to the cluster shapes for a crowdion,
parallel stacked crowdions, ring and two dumbbells arranged in T like structure.
The figures show three clusters for each of the shapes. The first cluster is
chosen randomly from the database and the other two are its closest neighbours
found using the chi-square distance from all the other features of clusters in
the database. The histograms when used to search similar clusters in the
database place qualitatively similar structures closer to each other than the
distinct ones.

The angle histograms for the similar shapes look qualitatively similar and 
for different shapes look different. In Fig. \ref{stacked-crowdion} the
cluster in (b) is oriented differently but still the histograms are closely
related. Similarly, in Fig. \ref{ring} neither the number of defects nor the
shapes perfectly match but still the feature representation retains the
qualitative similarity of the ring like structure.

In the Fig. \ref{linear} the histogram shows high occurrences of angles 0 and
180 degrees which is true for a linear crowdion cluster only. Similarly, other
shapes do show characteristic histograms.

\subsection{Dimensionality Reduction and Classification}

\begin{figure}[H]
  \centerline{\includegraphics[width=.9\linewidth]{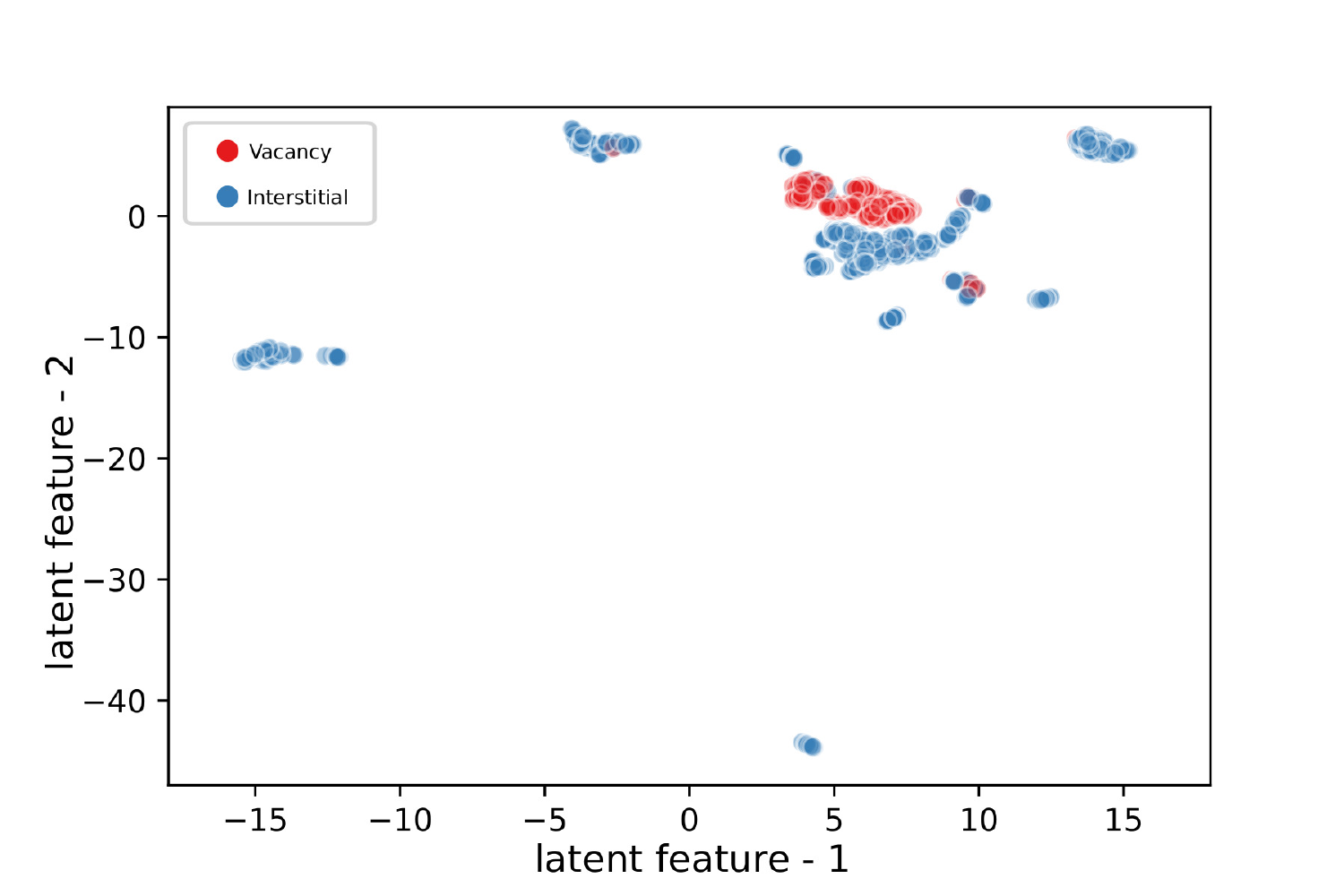}}
  \caption{\label{iorv}
    The UMAP embeddings of the cluster features. The vacancy and interstitial
    clusters are shown with different colors. The axes represent latent feature space of UMAP transformation.
}
\end{figure}

Although the features we found can be used to find similar structures, we need
dimensionality reduction to build 2D or 3D plots that can qualitatively show
the relationship between the different clusters. Fig. \ref{iorv} shows the
plot of angle and distance features of all the clusters in the database reduced
to two dimensional surface using UMAP. Each point in the plot represents a
cluster feature. The algorithm tries to maintain the local and global
relationship of the shape features. The figure shows that the algorithm embeds
vacancy clusters together, separate from interstitial clusters (the term
interstitial clusters here and in further discussions implies the clusters that
have more interstitials than vacancies). Broadly, the interstitial and vacancy
clusters do have different shapes. The embeddings found by the UMAP can change
with the change in the parameters that mostly govern the trade-off between
local and global structure and distribution of points in space. However, the
overall nature of the embeddings remain the same.  The embeddings of the t-SNE
are also similar with small differences in the details as we show later in this
section (Fig. \ref{classes-tsne}).

\begin{figure}[H]
  \centerline{\includegraphics[width=.9\linewidth]{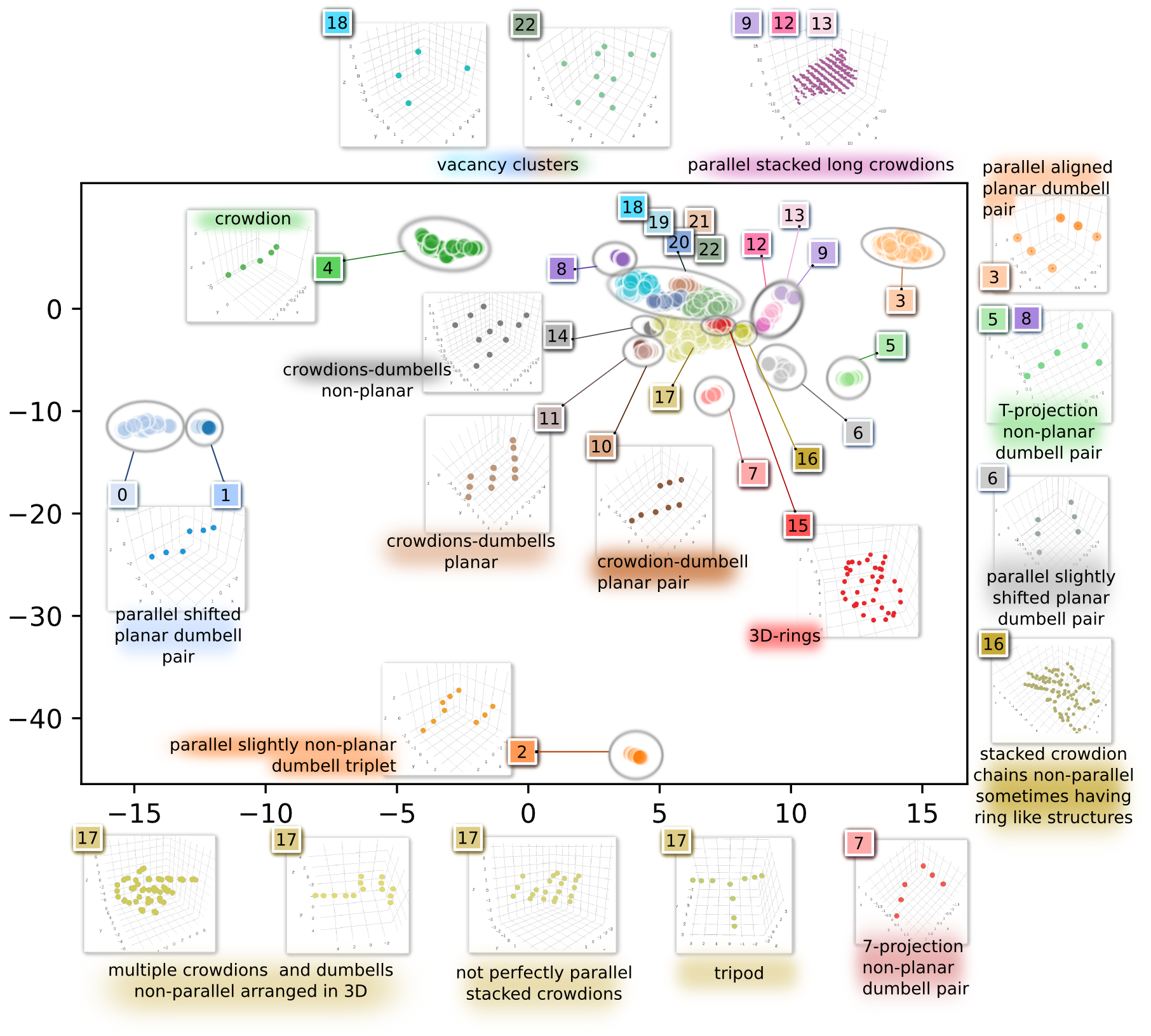}}
  \caption{\label{classes}
    Classification of the clusters using UMAP and HDBSCAN. Different classes
    are labelled from 0 to 22. The annotations show characteristic shapes for the
    classes. The axes represent latent feature space of UMAP transformation.
  }
\end{figure}

\begin{figure}[H]
  \centerline{\includegraphics[width=.9\linewidth]{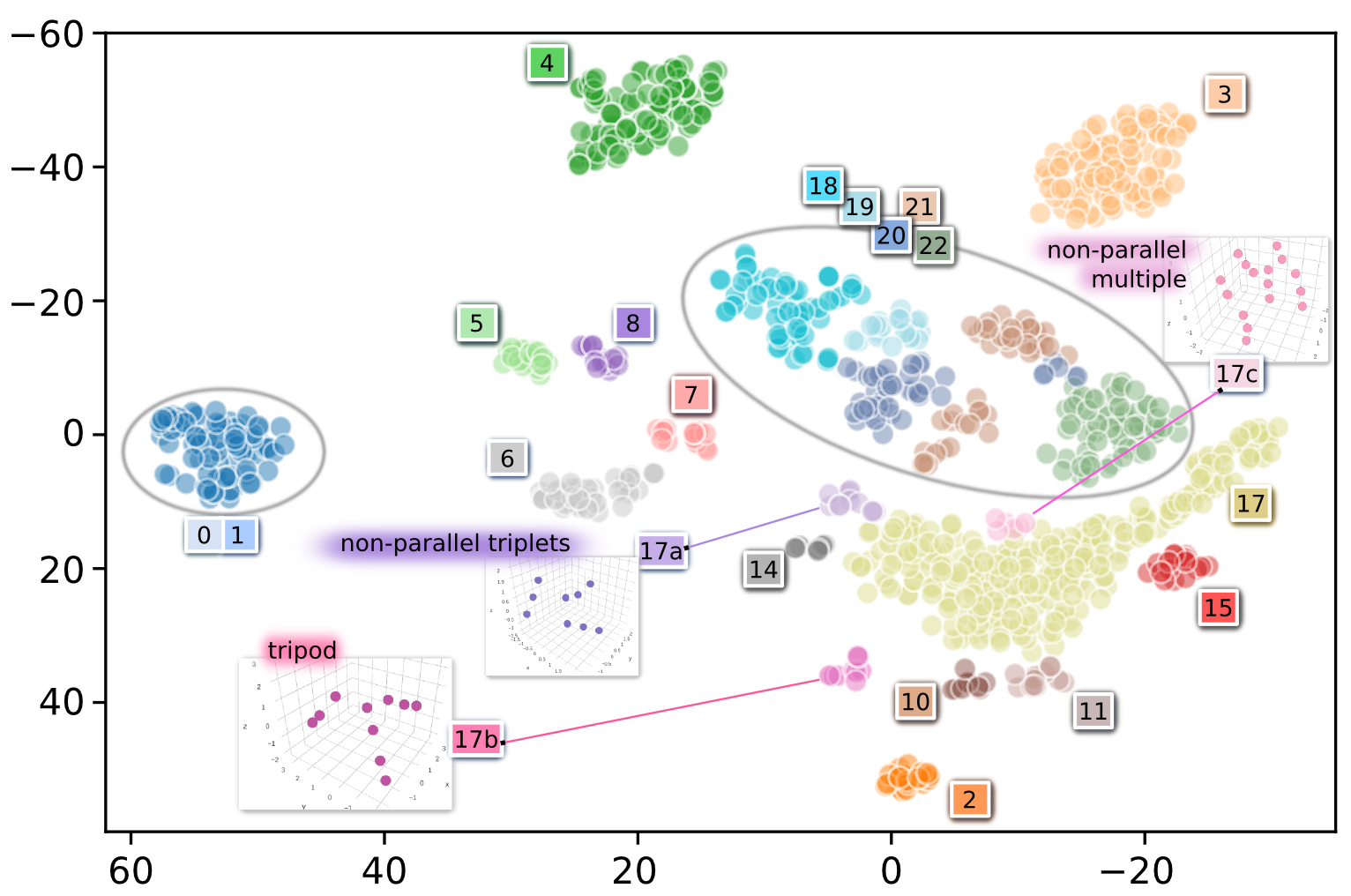}}
  \caption{\label{classes-tsne}
    Classification of the clusters using t-SNE and DBSCAN. The class labels
    are same as Fig. \ref{classes}. The annotations show characteristic shapes
    only for the classes that are new in this classification. The axes
    represent latent feature space of t-SNE transformation.
  }
\end{figure}

Fig. \ref{classes} shows the classification suggested by the HDBSCAN algorithm.
The classes are labelled from zero to twenty-two. The shape and name for each
class are also annotated in the figure. 

Each class represents shapes of exactly the same type, or closely resembling
shapes.  In a single class label, the structures can gradually change from one
end to the other, e.g. the left end of the 4th class represents perfectly
collinear crowdions, while as we move towards the right end (greater value of
latent-feature-1) the defects in the crowdions start to look slightly distorted
or not perfectly in-line with each other. These ends on zooming in might be
seen as sub-classes in a class and by using different values of parameters for
the unsupervised classification, might be separated into different classes. The
global relationship is also revealed when we move from one class to another.
The classes that are close in shape to each other appear close by, e.g. 15th
class represents ring-like structures while 16th class, appearing nearby,
represents big 3D shapes that have crowdions and dumbbells in random
orientations.  These can sometimes form partially ring like structures towards
the ends. Instead of complete random directions there can be two or more
parallel stack of crowdions with different orientations giving an overall
appearance of random orientations. Similarly, classes 9th, 13th and 12th show
stacked crowdions gradually decreasing in size going from right to left. The
trend of decrease in size continues in the 17th class. Indeed, the 17th class
has on the periphery structures that look like that of nearby classes and they
gradually fade from one structure to the other. The fading of one type of
structure into some partially recognizable shape and then finally into a
different structure is enticing to see as the structures change size,
dimensionality and slowly lose their qualitative characteristic identity. For
some of the classes which are totally separate from all others such as 2nd, 4th
etc., it is difficult to see the global relationship.

\begin{figure}[H]
  \centerline{\includegraphics[width=.9\linewidth]{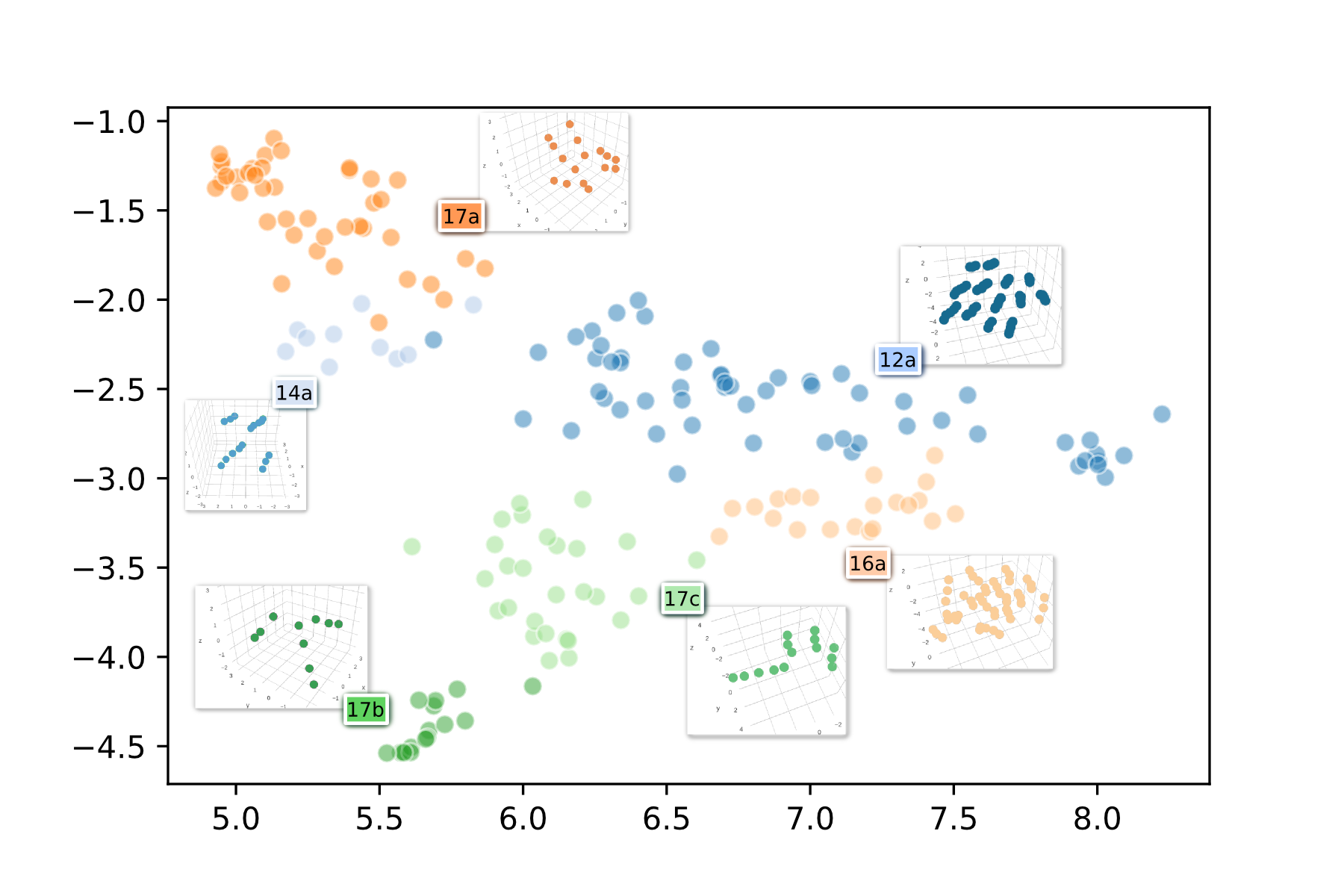}}
  \caption{\label{sub-classes}
    Sub-classification of the clusters in 17th class.
    The annotations show characteristic shapes for the classes. The axes
    represent latent feature space of UMAP transformation.
  }
\end{figure}

The density based algorithms classify only the similar density classes. Since
the structures in 17th class are evenly spaced due to their gradual change and
are relatively lesser in number, they are classified into one class. However,
if we zoom in, the density clusters can be seen. Fig. \ref{sub-classes} shows
the sub-classes of 17th class found by applying HDBSCAN to only the points in
the class. The sub-classes are named after the classes with which they most
relate to. The class 12a is a continuation of long parallel crowdions however
shorter than the 12th class. The reduction in size continues in 14a with four
or five shorter crowdion chains. The 14th class had only three number of
relatively shorter crowdions and dumbbells. The 16a class is just like 16th
class but with lesser number of defects and shorter crowdion chains. The
16a class fades into 17c with smaller and more planar randomly oriented
dumbbells and crowdions. Similar to 14a and 12a, 17a has three to five
crowdions and dumbbells but these are all randomly oriented in a 3D
arrangement unlike planar 17c. Sometimes the dumbbells or crowdions share
a vacancy to form a characteristic shape. 17b shows a distinct tripod
like arrangement of three dumbbells with shared extra defect at top. Sometimes
one of the leg of the tripod is missing an interstitial - vacancy pair at the
end.

The classification shown in Fig. \ref{classes} has a few misclassifications 
that include assignment of shapes into one class (label 17th) that
qualitatively can be divided into different classes and classification of
qualitatively similar shapes into two different classes as found in label 5th
and 8th, 0th and 1st. While the former case is improved by sub-classification
as discussed above, the latter is dealt by the DBSCAN classification on t-SNE
embeddings as shown in the Fig.  \ref{classes-tsne}. After accounting for these 
corrections we get 21 classes and 5 sub-classes making a total of 26 classes.

The t-SNE embeddings also
show some sub-classes (17a, 17b, 17c) of the 17th class. 
The overall shape of the embeddings is very similar to that
of UMAP in Fig. \ref{classes}, however the points are more evenly spaced. The
placement of different classes is also very similar e.g. the 4th class of
crowdions is at the top, 2nd class at the bottom, 17th class has similar shape
and above it are the vacancy classes. However, there are some differences as listed
below:

\begin{itemize}
\item Classes 0 and 1 that have qualitatively same shape are classified into
the same class unlike before.
\item Classes 5th and 8th that have qualitatively same shape are placed side by side.
\item There are more classes with distinct shapes around 17th class (17a, 17b, 17c) that
  were also found in the subclassification on UMAP embeddings.
\item Classes 9th, 12th and 13th that show gradation of sizes of big stacks of
crowdion chains are not classified separate from 17th class. Class 16th is also
missing.
\end{itemize}

\subsubsection{Properties of Classification}

We now look at the quantitative properties of the clusters present in the various
classes. The dimensionality of classes range from one-dimensional crowdions
(4th label) to highly planar arrangement of dumbbells and crowdions (0, 1, 3,
11) to 3-D complex structures (15, 16, 16a). The size of classes range from
single crowdion (4) to big stacked chains of defects (9) of 1/2(111) loops. The
clusters where net defect count gives vacancies are classified into 5 classes
from 18 to 22 while all other classes have clusters that have interstitial
nature after net defect count.

\begin{figure}[H]
  \centerline{\includegraphics[width=.9\linewidth]{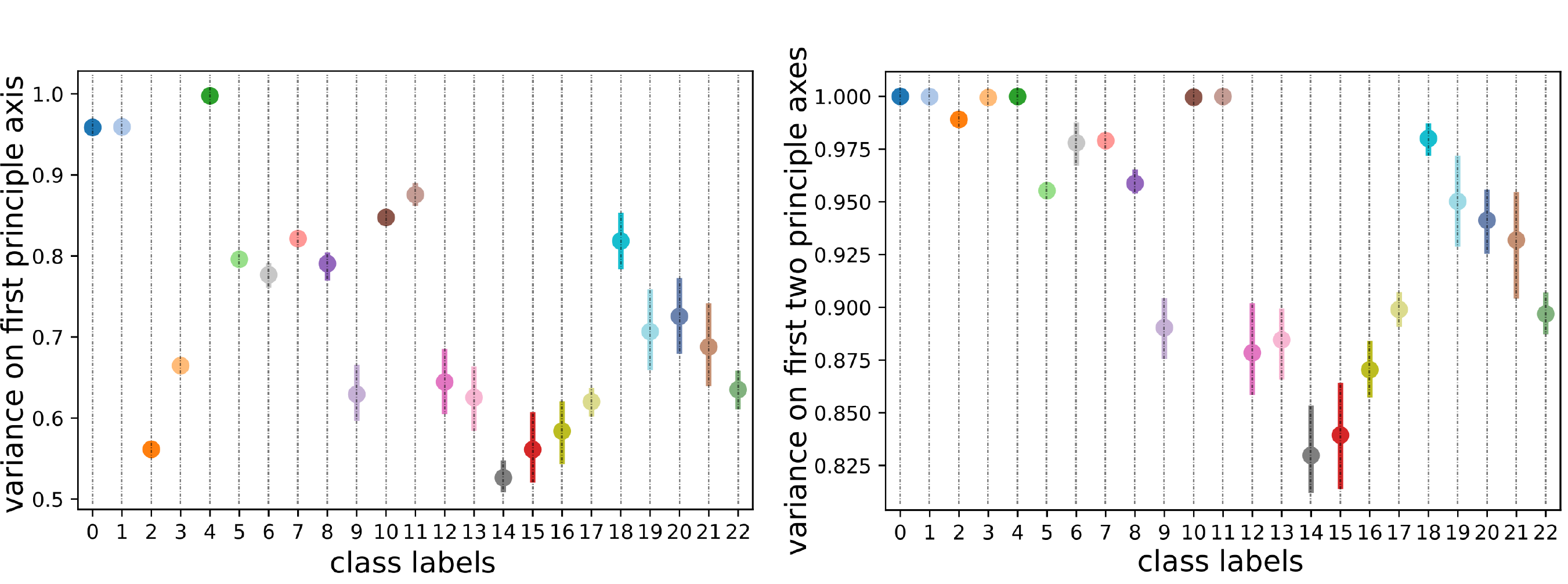}}
  \caption{\label{vr}
    The point estimates of variances as described by (a) first principle axis,
    (b) first two principle axes together, for different classes. The high
    value in first implies linearity and for low value in first, high value in
    second implies more planarity.
}
\end{figure}

Fig. \ref{vr} shows the dimensionality of the various classes, plotting the
point estimates of variances along the principle axes. The principle axes of a
cluster are found using the PCA (principle component analysis) using SVD
(singular value decomposition). The variance along an axis tells about the
spread along that axis. For a linear cluster, all the defects are spread along
the first principle axis and the variance along the first principle axis will
be 1.0. For a perfectly planar structure points would lie on only two principle
axes and the third axis will have no spread making the variance along the first
two principle axes 1.0. Thus, The high variance on first principle axis in
figure (a) implies a more linear shape. For a non-linear shape, a high variance
on first two principle axes, as shown in figure (b), implies a more planar
structure. It can be seen that one dimensional crowdions (label 4) are
perfectly linear while class labels 0, 1, 3, 10 and 11 are perfectly planar.
Classes 2, 6, 7 and 18 are almost planar while 14, 15 and 16 are distinctively
3D structures.

Among vacancy cluster classes we see that the 18th class is more planar than any
other vacancy cluster class also owing to its very less number of vacancies. The
planarity decreases gradually from 19 to 21 while class 22 can be said to be
most non-planar again owing to its distinctively big clusters.

\begin{figure}[H]
  \centerline{\includegraphics[width=.9\linewidth]{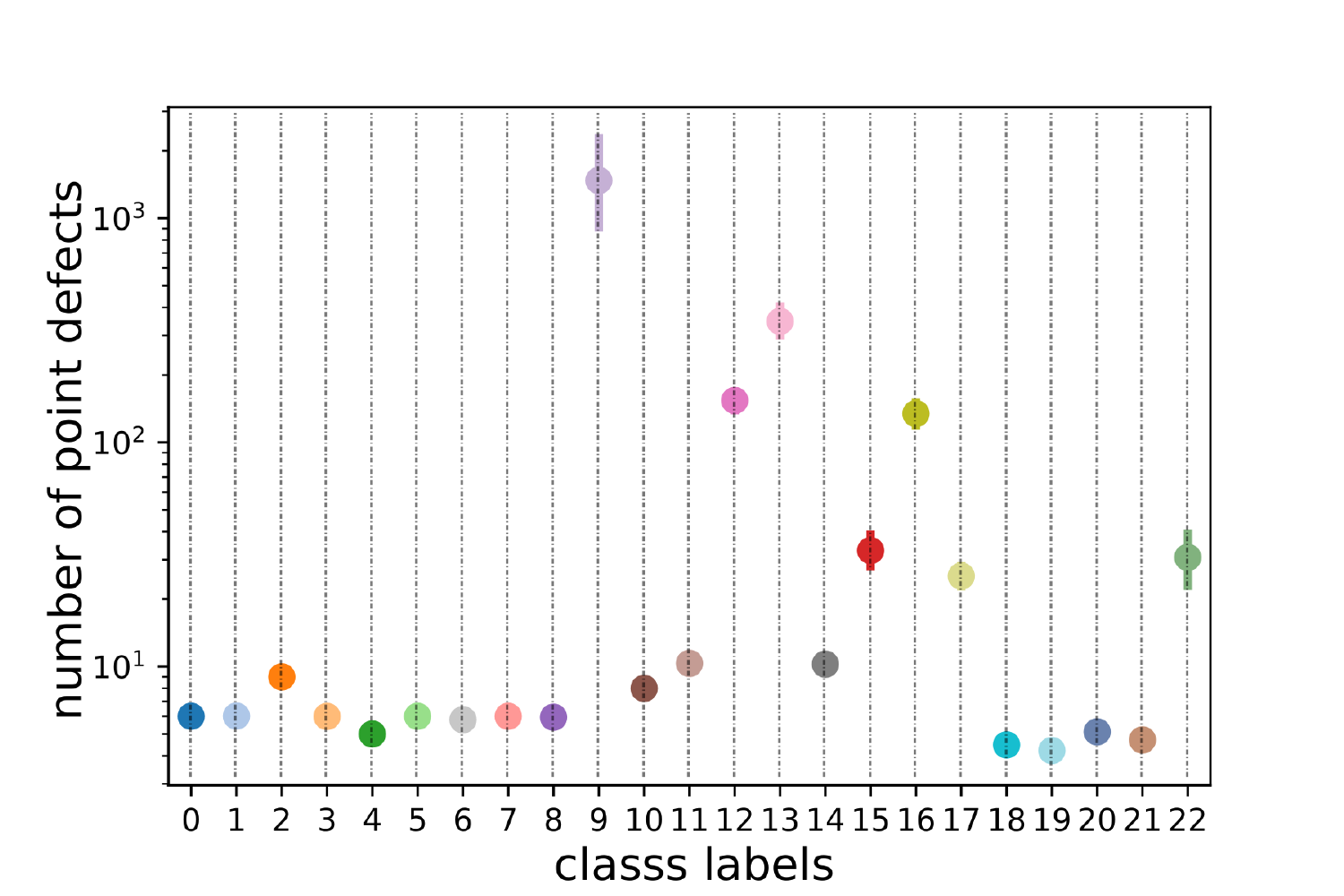}}
  \caption{\label{sz}
    The point estimates of number of total point defects in the clusters
    assigned to different classes.
}
\end{figure}

Fig. \ref{sz} shows the point estimates for total number of defects,
including extra vacancy-interstitial pairs, in clusters of each class. Most of
the classes have clusters of sizes ten or less while 9th class has really large
clusters followed by 13, 12 and 16. The biggest of the vacancy clusters are
classified in class 22.

\begin{figure}[H]
  \centerline{\includegraphics[width=.9\linewidth]{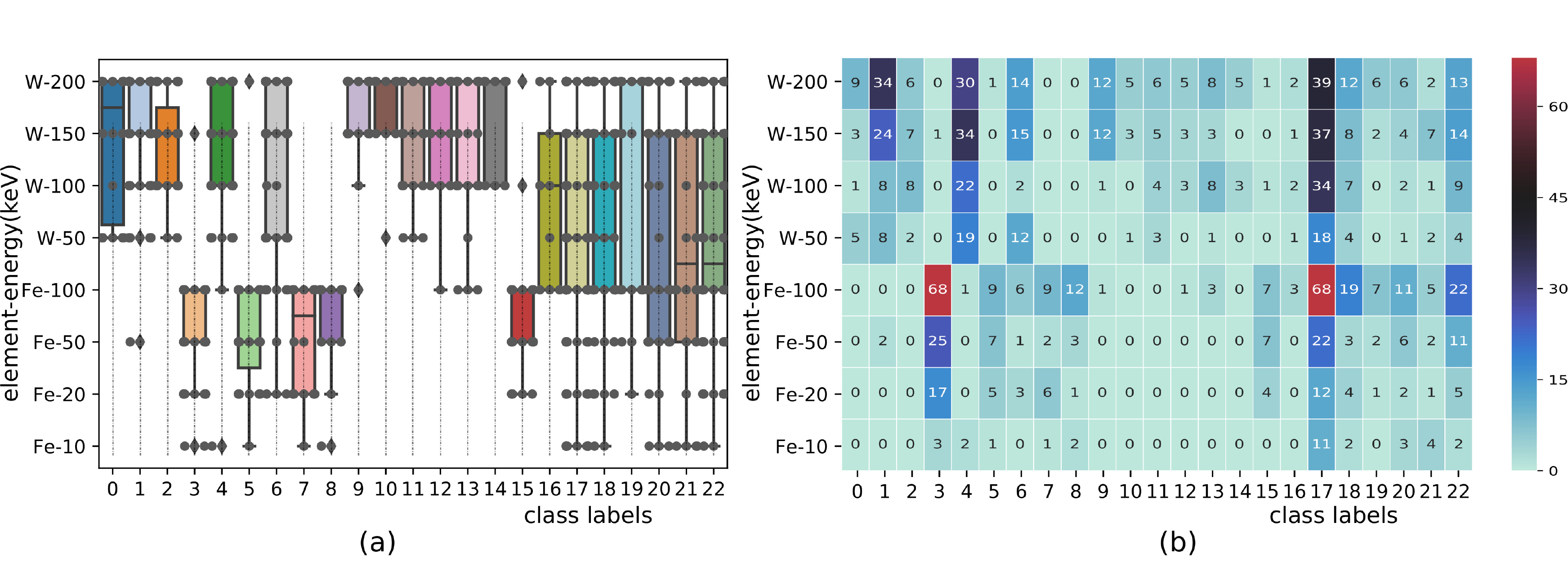}}
  \caption{\label{elemNrg}
    The distribution of clusters of different shapes among various elements and
    energies shown using (a) box-plot, (b) heat map.
}
\end{figure}

Fig. \ref{elemNrg} shows the distribution of clusters among classes and
cascades of both the elements at each energy of primary knock-on atom with the
help of (a) box-plot and (b) heat map with annotated values. Our analysis shows
the difference in interstitial cluster morphologies that arises as a result of
the different ground state configurations of the interstitial defect, which in
W adopts a $\langle111\rangle$ crowdion configuration (class 4) and in Fe a
$\langle110\rangle$ dumbbell configuration. In addition, the tendency in W to
form large defects \cite{andrea}, as opposed to the predominantly small defects
in Fe, is quantified in our results. The figure shows some shapes such as those
included in classes 3 or 5, 7, 8 and 15 that are exclusive to Fe, while classes
0 or 1, 2, 4, 9 to 14 and 16 that are predominantly present in W. The
classes exclusive to W have perfectly parallel stacked crowdion chains of
various sizes arranged in different fashion, while in Fe small dumbbells and
sometimes longer distorted crowdion like chains are arranged in random
orientations, except for class 3 that has a almost perfectly aligned pair of
parallel dumbbells. More tendency to form ring like shapes in Fe can also be
understood as an extension of this. Some of the classes of clusters obtained in
Fe have been reported in \cite{DEZERALD2014219}. The well investigated 3-D
combination of ring like structures in Fe viz. C15 resembles the class 15
structures classified by the unsupervised algorithm. However, a more convincing
study to match the different structures found in this work and the earlier
reported cluster shapes can be carried out in the future.

\begin{figure}[H]
  \centerline{\includegraphics[width=.9\linewidth]{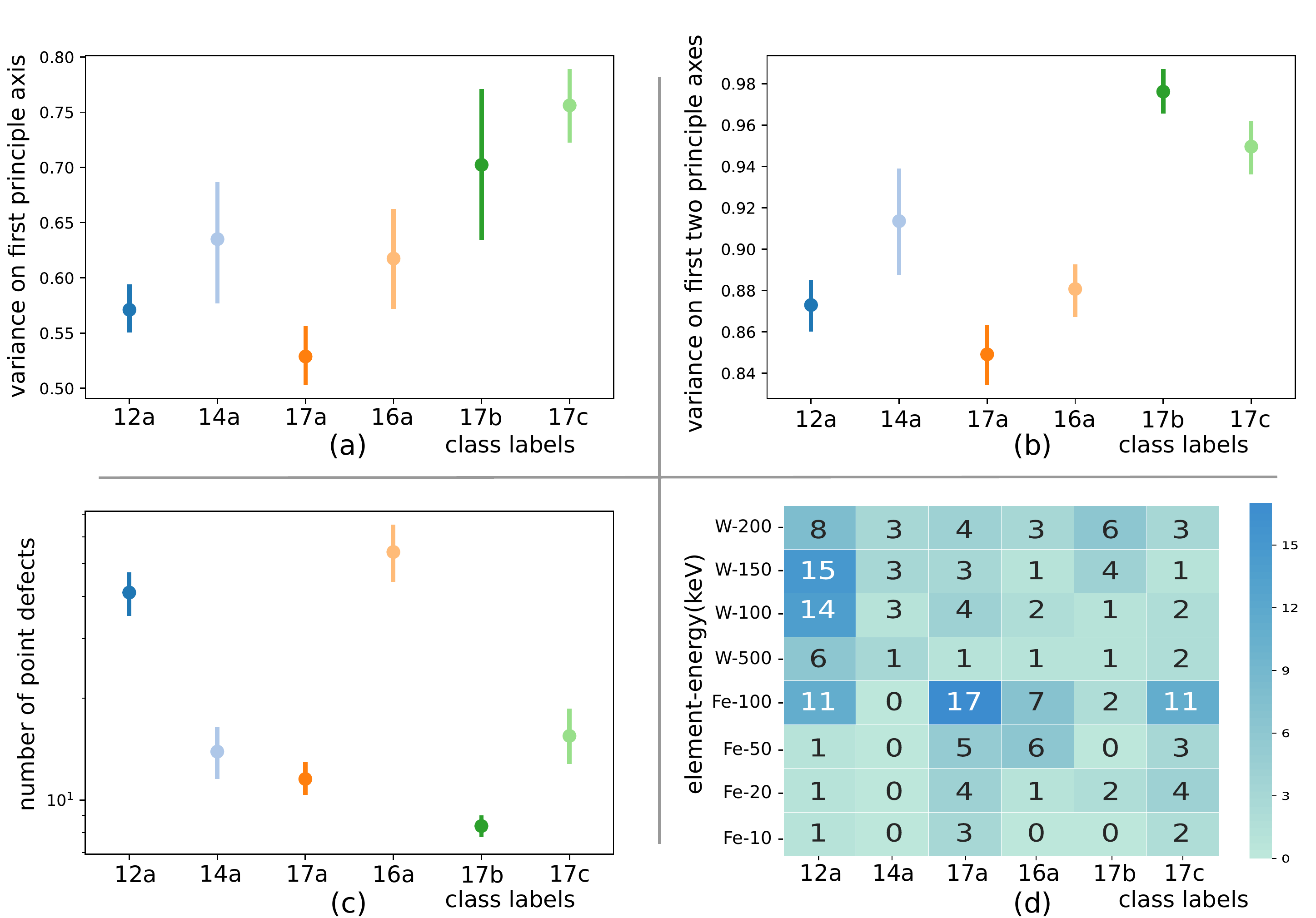}}
  \caption{\label{subclass-props}
    Showing the properties of various subclasses of 17th class. The figure
    (a) and (b) show the point estimates of variances as described by 
    first principle axis and first two principle axes together, respectively.
    The high value in first implies linearity and for low value in first, high
    value in second implies more planarity. The figure (c) shows
    the point estimates of number of total point defects in the clusters
    assigned to different classes. The heatmap in figure (d) shows distribution
    of various sub-classes among various elements and energies.
  }
\end{figure}

Fig. \ref{subclass-props} shows the dimensionality, sizes and distribution of
clusters in the sub-classes among the elements at each energy. The sizes and
dimensionality of the sub-classes is as suggested by their qualitative shape
17b and 17c are almost planar, the size of 12a is less than 12th class and
more than 14a which itself has more defects than the 14th class. 

More interestingly,
the classes having parallel stacked crowdions and dumbbells viz. 12a, 14a again
show up more in W while Fe clusters are predominantly formed of randomly
oriented dumbbells appearing in classes viz. 17a, 17c and 16a. However, Fe
clusters do appear in classes such as 12a with groups of long but slightly
distorted crowdion like chains and W clusters appear in classes like 17a and
17c with randomly oriented distorted (non collinear) dumbbells. The 17b shows a new
tripod like arrangement of dumbbells almost exclusive to W.

Broadly the observations and corresponding descriptions of the class labels
can be summarized as follows:

\begin{enumerate}
  \item Linear - specific to W:
  \begin{enumerate}
    \item Label 4: Crowdions (size 5): Mostly perfectly collinear but sometimes a bit off.
  \end{enumerate}
  \item Planar:
  \begin{enumerate}
    \item Dumbbell pairs:
      \begin{enumerate}
        \item Labels 0, 1: Parallel shifted - specific to W
        \item Label 3: Parallel aligned - specific to Fe
        \item Label 6: Parallel slightly shifted - more common in W
      \end{enumerate}
    \item Crowdion-Dumbbells - specific to W:
      \begin{enumerate}
        \item Label 10: Pair
        \item Label 11: Multiple
      \end{enumerate}
    \item Vacancy clusters: Label 18
  \end{enumerate}
  \item Small, slightly non-planar structures:
    \begin{itemize}
      \item Label 2: parallel dumbbell triplet - specific to W
      \item Labels 5, 8: T-projection dumbbell pair - specific to Fe
      \item Label 7: 7-projection dumbbell pair - specific to Fe. This class also has some twisted dumbbell pairs that only partially resemble  the structure shown in Fig. \ref{classes}.
      \item Label 17b: crowdion dumbbell triplets - specific to W
      \item Label 17c: multiple crowdions and dumbbells - more in Fe 
      \item Labels 19, 20, 21 Vacancy clusters.
    \end{itemize}
  \item Small 3D structures
    \begin{itemize}
      \item Label 14, 14a: parallel crowdions-dumbbells - specific to W.
      \item Label 15: Ring like 3D-shapes - More common in Fe.
      \item Label 17a: small number (3 to 5) of randomly oriented dumbbells and
    crowdions, more in Fe.
    \end{itemize}
  \item Bigger 3D shapes - more common in high energies
    \begin{itemize}
      \item Label 16, 16a: dumbbells and crowdion like chains in different orientations
        sometimes having ring like arrangements at the ends. 
      \item Labels 9, 12, 12a, 13: parallel stacked crowdion chains - 9 (biggest), 13 (bigger), 12 (smaller), 12a (smallest).
      \item Label 
     \item Label 22: Big vacancy clusters.
    \end{itemize}
\end{enumerate}

The results shown can be used to characterize and study the class of clusters
in new data of collision cascades. The UMAP or t-SNE embeddings can be found for
a new defect cluster and then the cluster can be assigned to one of the already
found classes based on its proximity. We can also find the statistical
distribution of occurrence of classes of defects produced for any element in an
energy range from the collision cascades simulation data which can then be used
along with the properties like diffusion profile, recombination criteria,
stability etc. for the shapes, in higher scale models like Monte Carlo methods,
production bias model etc. These properties are known for some of the shapes while
for others these need to be explored.


\section{Conclusion}

We have described a method to efficiently process and analyze the structures of
clusters from MD simulations of collision cascades. We have discussed efficient
algorithms starting from identification and clustering of defects, to feature
engineering for cluster shapes, to their classification and visualization. We
use well established approximate algorithms from machine learning that enable a
solution with minimum inputs or assumptions and high robustness to noise. The
results can be used to study and add shape and structure based information of
defect clusters to higher scale models of radiation damage.

We have applied our methods and discussed elaborate results for collision
cascades in Fe and W for a wide range of PKA energies. We show that the
features we introduce work well for the pattern matching of defect clusters.
The dimensionality reduction algorithm helps visualize the relationships
between the clusters such as type of clusters (interstitial or vacancy), size
of the cluster and dimensionality of the cluster. The unsupervised
classification algorithms give twenty six classes that we identify by
different qualitative names based on the shapes they represent. The
classification can be used to quickly study the kind of clusters produced in
new elements and energy ranges. The classification and taxonomy also helps to
systematically study and explore the nature of defects caused by collision
cascades in different elements and energy ranges and use the properties for
that shape in higher scale models in a multi-scale radiation damage study. A
next step in a multi-scale model can be to assign the properties such as
diffusivity, recombination, thermal stability etc. to the identified classes.
While for some cluster structures found, the diffusion profile (sessile or
glissile, dimensionality of diffusion, diffusion coefficient etc.) is known,
for others it needs to be examined. The interaction of the clusters with other
defects and grain boundaries affects the micro-structural changes due to
irradiation.

We have laid out the steps that lead to a good classification and study of
clusters of defects with the use of state-of-the-art algorithms. Moreover, it
also opens up the scope of exploration of different specific algorithms for
particular steps such as the use of other features like shape barcodes from
topology, use of other clustering algorithms like K-Means etc. The methods can
also be further extended to study shapes of subcascades and even full collision
cascades.

\section*{
  Acknowledgements
}
This work was inspired by the IAEA Challenge on Materials for Fusion-2018. AES
acknowledges support from the Academy of Finland through project No. 311472.

\bibliography{content}

\end{document}

%% file: structure.tex
\usepackage[
nochapters, 
beramono, 
eulermath,
pdfspacing, 
dottedtoc 
]{classicthesis} 

\usepackage{arsclassica} 

\usepackage[T1]{fontenc} 

\usepackage[utf8]{inputenc} 

\usepackage{graphicx} 
\graphicspath{{Figures/}} 

\usepackage{enumitem} 

\usepackage{lipsum} 

\usepackage{subfig} 

\usepackage{amsmath,amssymb,amsthm} 

\usepackage{varioref} 

\theoremstyle{definition} 

\theoremstyle{plain} 

\theoremstyle{remark} 

\hypersetup{
colorlinks=true, breaklinks=true, bookmarks=true,bookmarksnumbered,
urlcolor=webbrown, linkcolor=RoyalBlue, citecolor=webgreen, 
pdftitle={}, 
pdfauthor={\textcopyright}, 
pdfsubject={}, 
pdfkeywords={}, 
pdfcreator={pdfLaTeX}, 
pdfproducer={LaTeX with hyperref and ClassicThesis} 
}